\documentclass{egpubl}
\usepackage[utf8]{inputenc}
\usepackage{graphicx}

\makeatletter

\providecommand{\tabularnewline}{\\}

\usepackage{wait2021}

\ConferencePaper      

\title{Creating good quality meshes from smooth implicit surfaces}

\author[Á. Sipos et al]
       {Ágoston Sipos and
        Péter Salvi
        \\
        Budapest University of Technology and Economics
       }

\PrintedOrElectronic

\usepackage{t1enc}
\usepackage{dfadobe}

        \let\tt=\ttfamily

\makeatletter
\let\ftype@table\ftype@figure
\makeatother

\makeatother

\begin{document}
\maketitle
\begin{abstract}
Visualization of implicit surfaces is an actively researched topic.
While raytracing can produce high quality images, it is not well suited
for creating a quick preview of the surface. Indirect algorithms (e.g.
Marching Cubes) create an easily renderable triangle mesh, but the
result is often not sufficiently well-structured for a good approximation
of differential surface quantities (normals, curvatures, etc.). Post-processing
methods usually have a considerable computational overhead, and high
quality is not guaranteed. We propose a tessellation algorithm to
create nearly isotropic meshes, using multi-sided implicit surfaces.
\end{abstract}

\section{Introduction}

Implicit surfaces are one of the main kinds of mathematical representations
used in geometric modelling. Implicit surfaces facilitate many operations,
but others - like direct visualization - are less efficient. For this
reason, when creating quick interactive previews, often indirect methods
like Marching Cubes\cite{Lorensen:1987,Newman:2006}  and variants
are used to create meshes which are then efficiently rendered.

\begin{figure}[b]
\includegraphics[viewport=0bp 0bp 700bp 700bp,width=0.48\columnwidth]{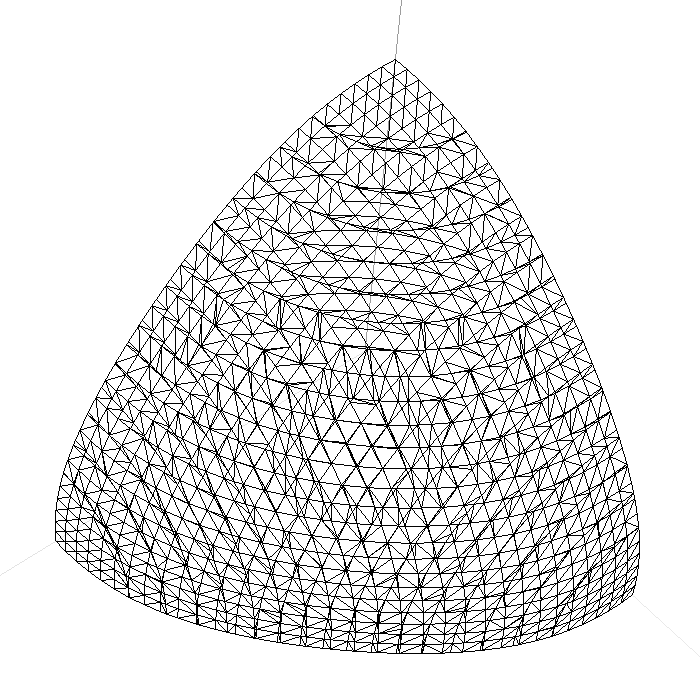}\includegraphics[viewport=0bp 0bp 700bp 700bp,width=0.48\columnwidth]{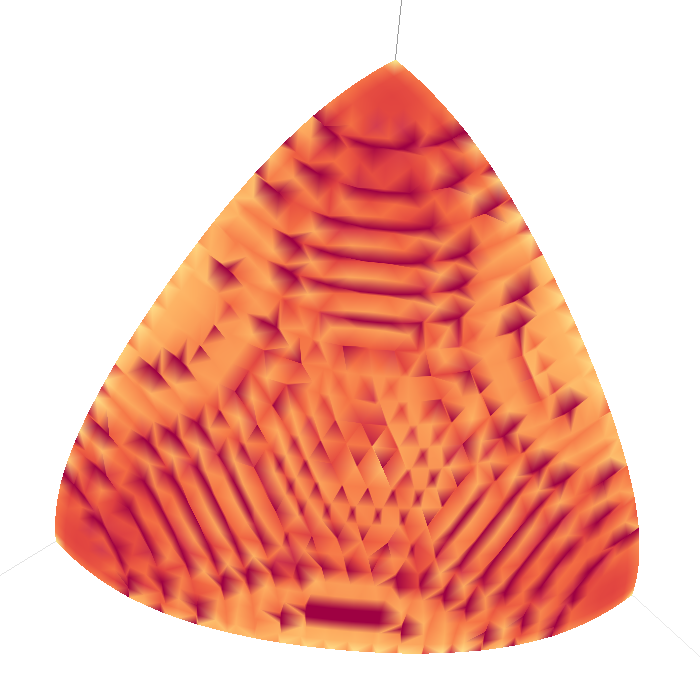}

\caption{Wireframe and approximated curvature map of a Marching Cubes mesh.}

\label{fig:mc}
\end{figure}

Such meshes, however, often do not have good enough quality to accurately
approximate differential surface quantities. In surface analysis,
it is often important to use coloring based on normal vectors, various
curvatures and other measures to identify potential defects. A mesh
directly produced by Marching Cubes (see e.g. Figure \ref{fig:mc})
is not useful for that purpose.

\begin{figure}[b]
\includegraphics[width=0.48\columnwidth]{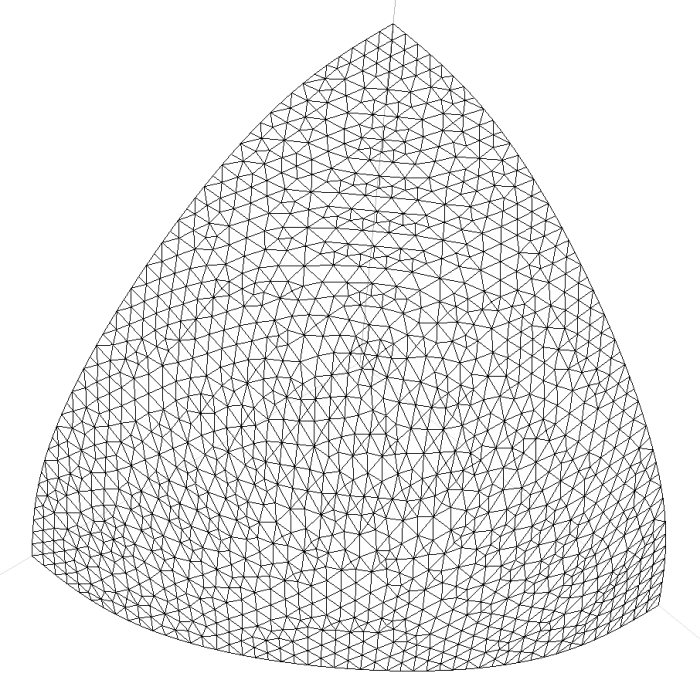}\includegraphics[width=0.48\columnwidth]{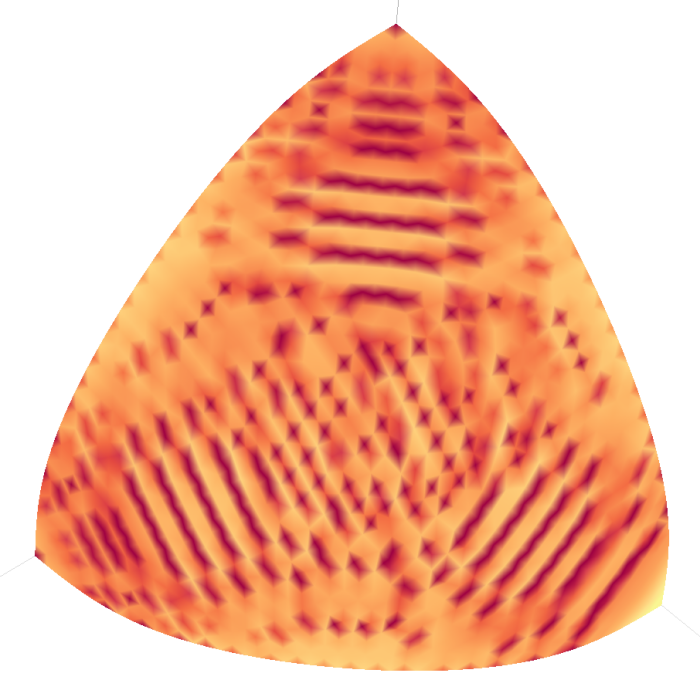}

\caption{Wireframe and approximated curvature map of a Marching Cubes mesh
after fairing.}

\label{fig:mc-lapl}
\end{figure}

Fairing and edge transformation methods\cite{Dietrich:2008} and algorithms
based on different principles\cite{deAraujo:2015} have been proposed
in the literature. However, these often have a high computational
overhead, many of them are complicated to implement, and they do not
guarantee that the resulting mesh will be free of anisotropic triangles
and vertices with unusually high or low valencies. For example, in
Figure \ref{fig:mc-lapl}, the mesh was processed by Laplace fairing,
which has considerably improved the quality of individual triangles,
but the mesh still includes degree 4 or degree 8 nodes where calculating
mean curvature becomes inaccurate.

Therefore, our approach is to first create a well-structured mesh
topology loosely around the surface, and then project its points onto
the implicit surface. In Section \ref{sec:Meshing} we will outline
our method, and in Section \ref{sec:Examples} surface examples will
be shown. An assessment of the algorithm will conclude the paper in
Section \ref{sec:Conclusion}.

\section{Meshing\label{sec:Meshing}}

Our method works with a space partitioning that divides the implicit
isosurface into several multi-sided pieces each containing only a
single surface sheet. One such structure is an octree, which we subdivide
while there are multiple surface sheets in a cell. However, the algorithm
can be used with any convex partitioning which is made of planar cuts
(an example will be shown in Subsection \ref{subsec:General-N-sided-patches}).

First we can create the corner points of the multi-sided surfaces,
which will be the basis of the algorithm. For each edge of a cell,
if its endpoints are on different sides of the surface, we need to
find the intersection point which will be the corner point. Then for
each face of the cell, we connect the intersections with an edge,
so that we have a boundary loop of the patch.

\subsection{Base mesh\label{subsec:Base-mesh}}

Next, a base mesh is created, using an auxiliary parametric surface,
which interpolates the corner points and the boundaries. The simplest
method is to use a generalized barycentric\cite{Floater:2015} combination
of the corner points over a regular $n$-sided domain:
\begin{equation}
p(u,v)=\sum_{i=1}^{n}\lambda_{i}\mathbf{p}_{i},\label{eq:bari}
\end{equation}
where $\sum_{i=1}^{n}\lambda_{i}=1$, and $[u,v]=\sum_{i=1}^{n}\lambda_{i}\cdot[u_{i},v_{i}]$.
$[u_{i},v_{i}]$ are the domain coordinates of the corners of the
regular $n$-gon; $\mathbf{p}_{i}$ are the corner points of the surface.
This ensures that the edges of the patch will run in the appropriate
face dividing the cell from its neighbour.

Alternatively, as the implicit surface can be highly curved, a simple
transfinite parametric surface can be used, like the multi-sided $C^{0}$
Coons patch\cite{Salvi:2020:WAIT} which is defined by its boundary
curves. If those do not have a simple parametric representation, they
can be given as dense polylines\footnote{As the mesh vertices will be projected onto the surface, the slight
error arising from the inexact curves will not meaningfully affect
the final result.} evaluated via tessellating each implicit boundary curve, which is
the intersection of the planar bounding face with the implicit isosurface.

\begin{figure}
\hspace{0.1\columnwidth}\includegraphics[width=0.8\columnwidth]{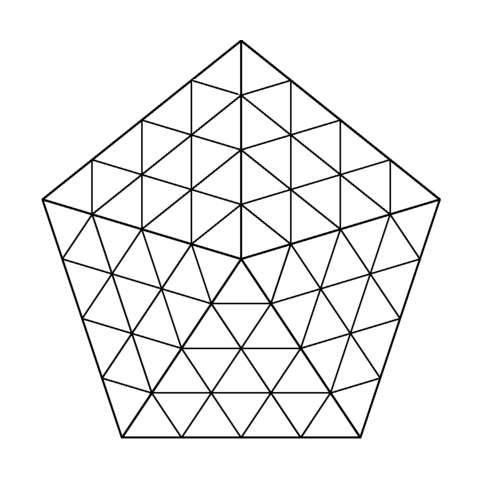}

\caption{Triangulation of a pentagon.}

\label{fig:triangulation}
\end{figure}

These surfaces can then be elegantly tessellated by dividing the $n$-sided
polygonal domain into triangles like in Figure \ref{fig:triangulation}.

\subsection{Projection}

When projecting the points of the mesh onto the isosurface, we must
ensure continuous connection to neighbouring patches. For that, we
need the points on the edge of the patch to remain on the face of
the cell after projection. We also want triangles to change their
size relative to each other as few as possible.

\begin{figure}
\includegraphics[width=1\columnwidth]{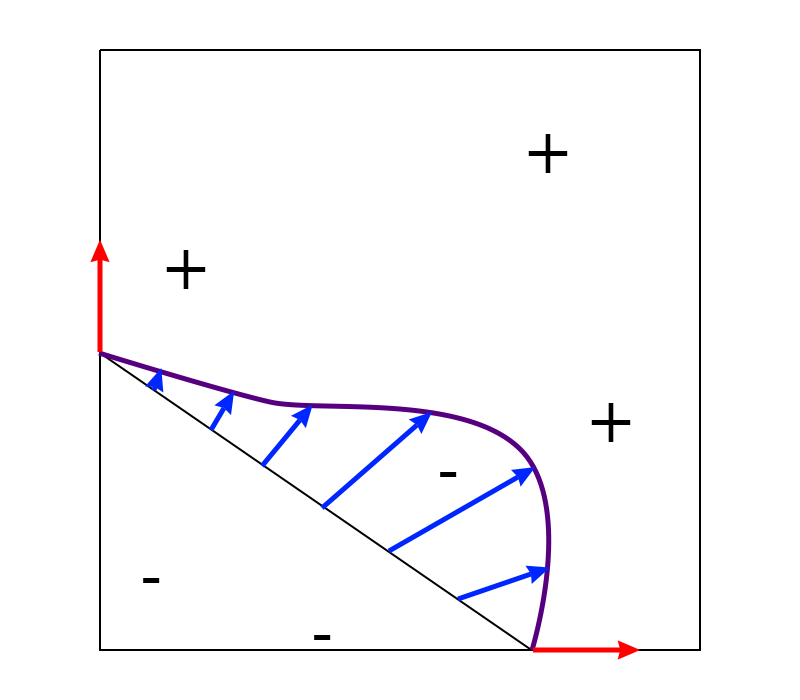}

\caption{2D example of projection direction interpolation.\protect \\
Purple: implicit surface, red: prescribed directions, blue: interpolated
directions. Signs denote the sign of the implicit function in that
region.}

\label{fig:projdir}
\end{figure}

We achieve this by defining a \emph{projection direction} (a 3-dimensional
vector) attribute for each vertex which will prescribe the line, along
which the point will be projected. The projection direction is prescribed
in the corners of the patch, and is then interpolated along the mesh
using barycentric coordinates. In the corners, they need to point
in the same direction from the isosurface (e.g. into the positive
half-space). See 2-dimensional example in Figure \ref{fig:projdir}.

If the boundary points have to be moved (i.e. we are not using a base
patch with exact boundary points), then at corner points the direction
shall be set as the direction of the edge of the cell there. This
ensures that in edge points the direction is inside the bounding plane,
as it is the weighted sum of two edges in that plane (all other barycentric
coordinates are zeros there). Thus, the projected points will also
remain there.

\begin{figure}
\includegraphics[width=1\columnwidth]{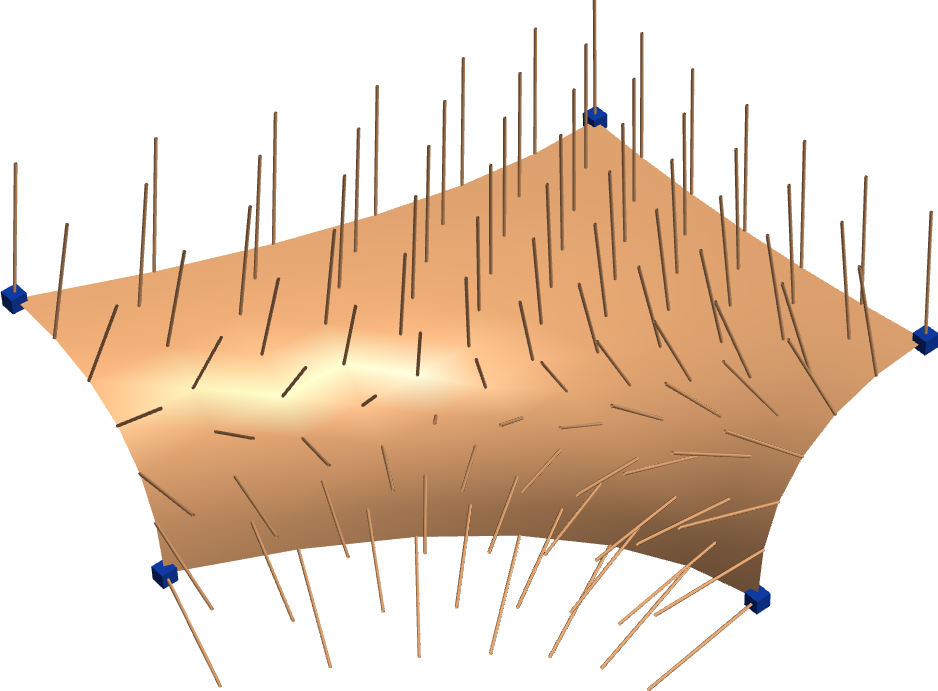}

\caption{Projection directions visualized on the surface of a $C^{0}$ Coons
patch.}

\label{fig:rays}
\end{figure}

See Figure \ref{fig:rays} for a visualization of these direction
vectors.

Now we can use ray marching to find the isosurface point for each
vertex. By checking the sign of the implicit function in the starting
point, we know in which direction the isosurface lays, relative to
the interpolated direction. In the end, we replace the mesh positions
with the projected positions, getting a mesh representing the piece
of the original isosurface inside the cell. Corner points are not
moved, and if the boundary curves were already approximated, boundary
points stay in place as well.

The above process unfortunately does not guarantee that correct results
are obtained. If the isosurface has high curvature variation or large
shape artifacts, the resulting mesh might contain abrupt jumps or
overlapping triangles. We prevent this by checking if the angle between
the ray and the gradient of the surface remains under a prescribed
threshold. If not, the resulting mesh is rejected, and we can try
to solve the problem by subdividing the cell. This typically indicates,
however, that the quality of the implicit surface would not be acceptable.

\section{Examples\label{sec:Examples}}

\subsection{Cell patches}

\begin{figure}
\includegraphics[width=0.5\columnwidth]{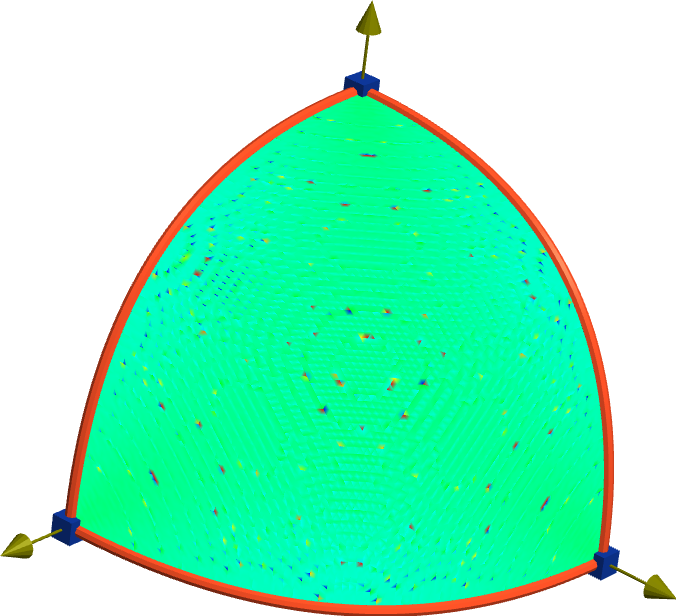}\includegraphics[width=0.5\columnwidth]{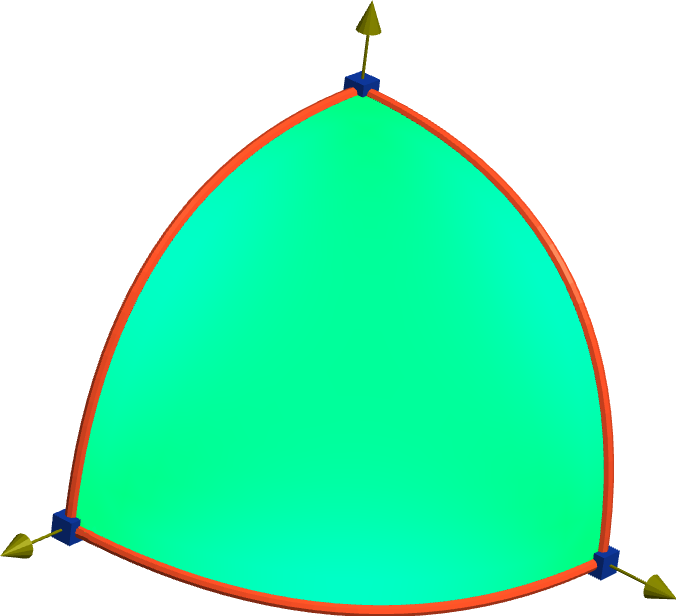}

\caption{Mean curvature map of a Marching Cubes mesh and by our approach (3-sided
surface in a cube cell).}

\label{fig:3sided}
\end{figure}

\begin{figure}
\includegraphics[width=0.48\columnwidth]{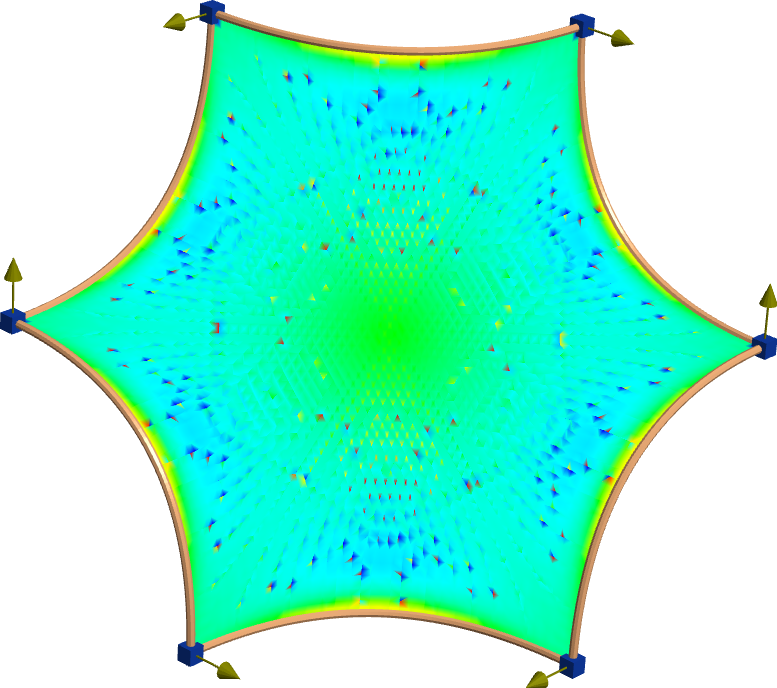}\hspace{0.04\columnwidth}\includegraphics[width=0.48\columnwidth]{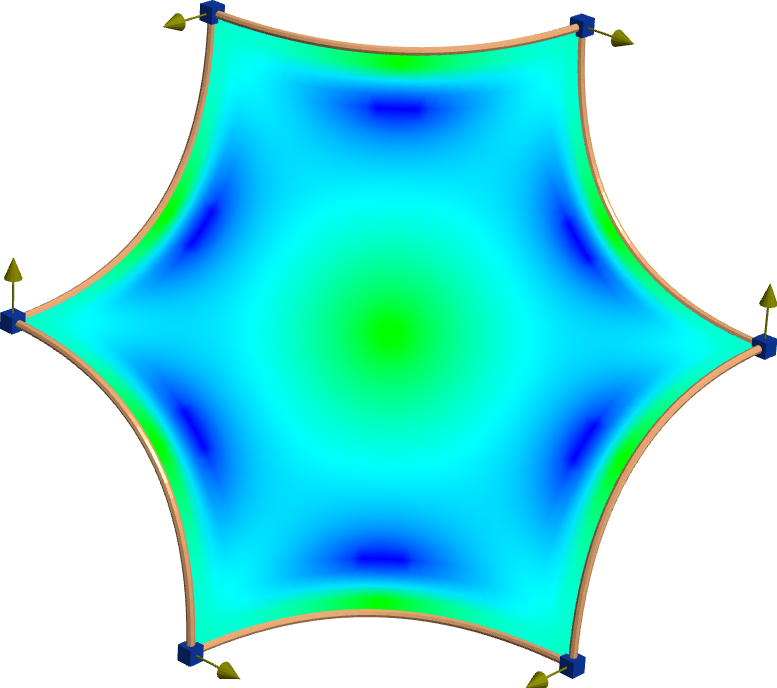}

\vspace{0.04\columnwidth}

\includegraphics[width=0.48\columnwidth]{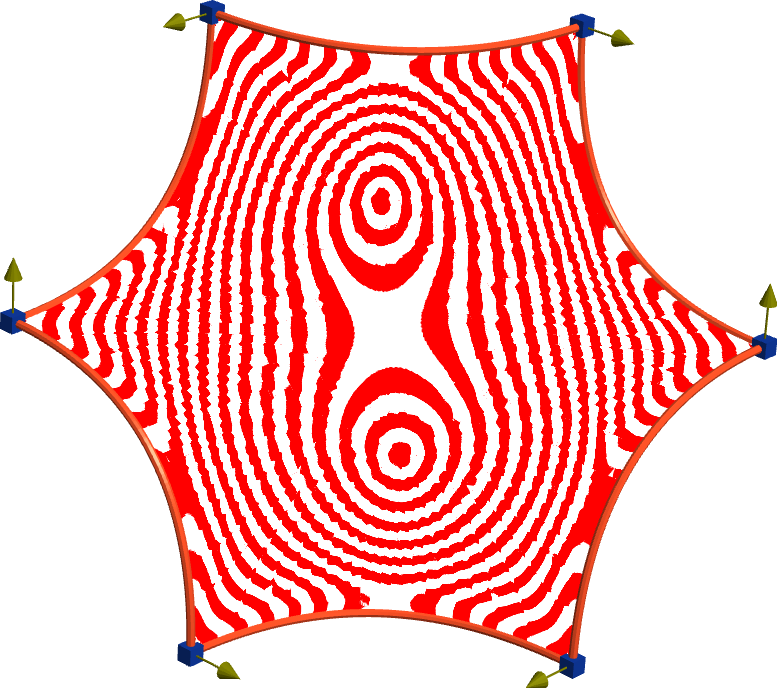}\hspace{0.04\columnwidth}\includegraphics[width=0.48\columnwidth]{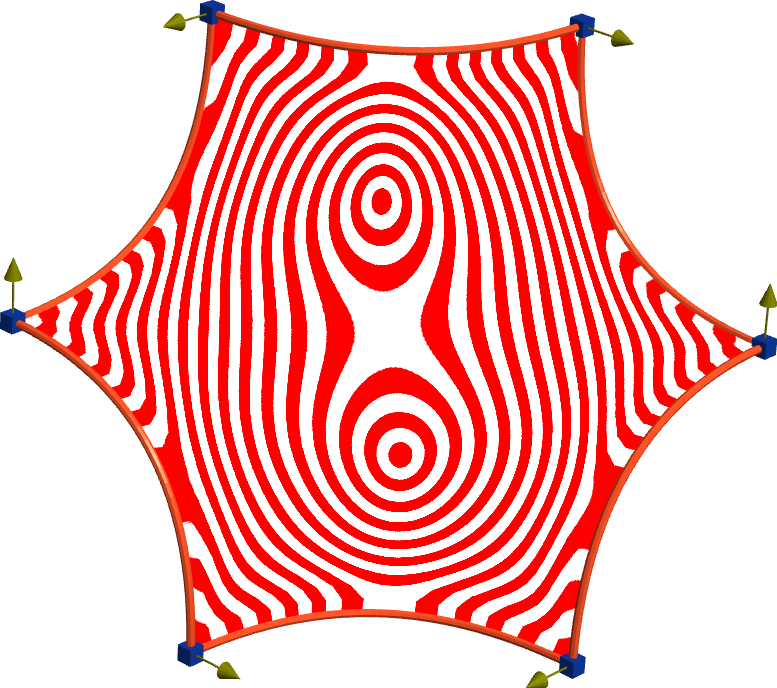}

\caption{Comparing a Marching Cubes mesh to our approach (6-sided patch). Top
row: mean curvature, bottom row: isophote lines.}

\label{fig:6sided}
\end{figure}

\begin{figure}
\includegraphics[width=0.48\columnwidth]{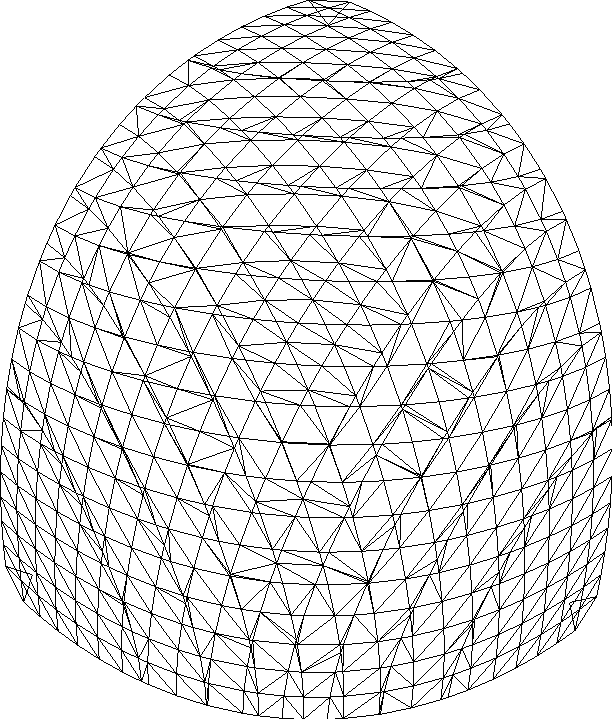}\hspace{0.04\columnwidth}\includegraphics[width=0.48\columnwidth]{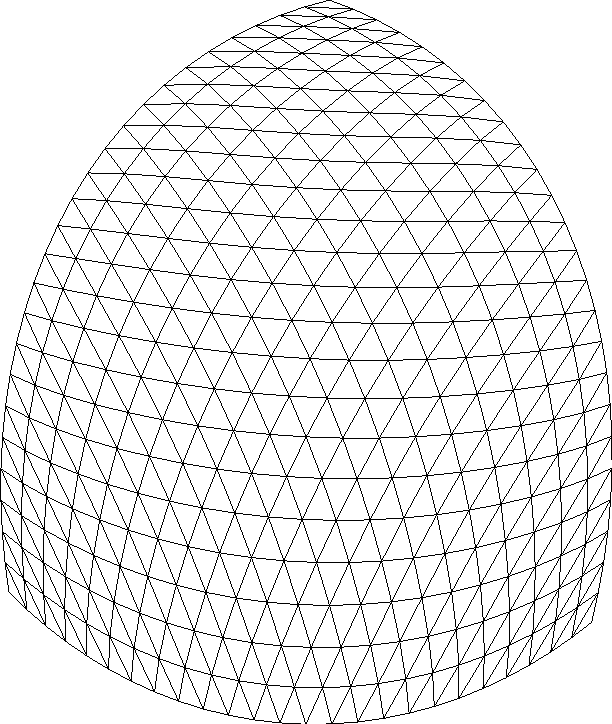}

\caption{Wireframe visualization of a Marching Cubes mesh (left) and a projected
mesh (right). See top row in Table~\ref{tab:results} for their details. }

\label{fig:wireframe}
\end{figure}

Surface patches inside cubic cells can easily be compared, as these
can be efficiently tessellated with both Marching Cubes and the proposed
algorithm, and they will represent the exact same surface. We show
such comparisons in Figures \ref{fig:3sided}, \ref{fig:6sided} and
\ref{fig:wireframe}.

\begin{table}
\centering%
\begin{tabular}{|c|c|c|c|}
\hline 
\multicolumn{2}{|c|}{Marching cubes} & \multicolumn{2}{c|}{Projection}\tabularnewline
\hline 
\hline 
Vertices & Time & Vertices & Time\tabularnewline
\hline 
688 & 20ms & 351 & 4ms\tabularnewline
\hline 
1760 & 58ms & 561 & 8ms\tabularnewline
\hline 
3881 & 136ms & 1326 & 14ms\tabularnewline
\hline 
\end{tabular}

\caption{Efficiency of generating meshes with various density from the surface
in Figure \ref{fig:wireframe}.}

\label{tab:results}
\end{table}

\subsection{General $n$-sided patches\label{subsec:General-N-sided-patches}}

\begin{figure}
\includegraphics[width=1\columnwidth]{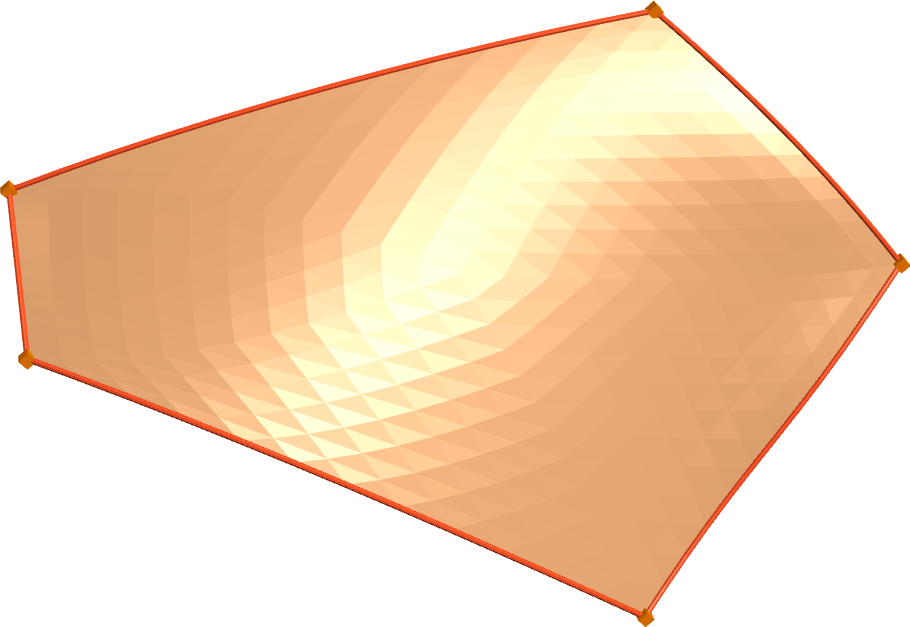}

\caption{Five-sided implicit patch bounded by general configuration planes.}

\label{fig:patch}
\end{figure}

In case of general space partitions bounded by planes, the resulting
surface can also be nicely triangulated, as in Figure~\ref{fig:patch}.

\subsubsection*{Approach to I-patches defined by control polyhedra}

\begin{figure}
\includegraphics[width=0.5\columnwidth]{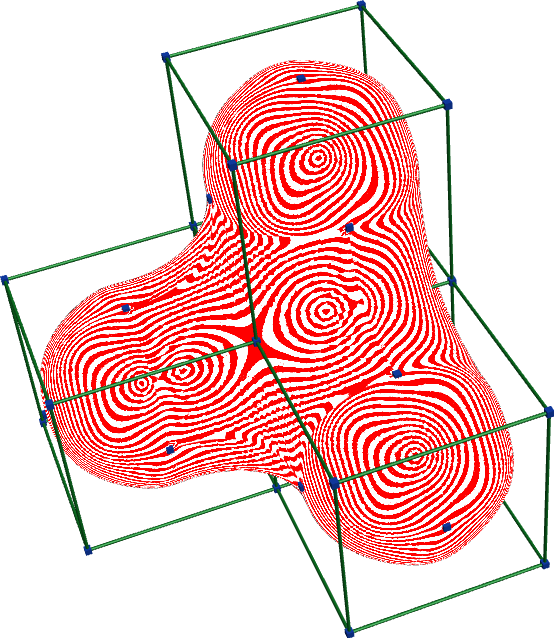}\includegraphics[width=0.5\columnwidth]{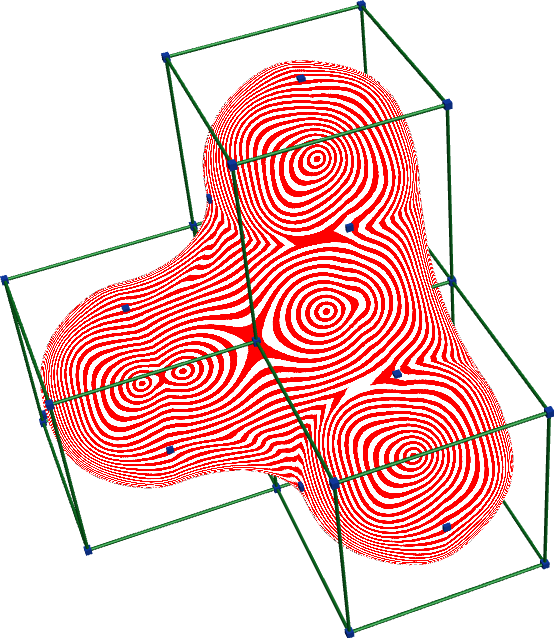}

\caption{Composite surface built from I-patches triangulated with Marching
Cubes (left) and projection (right).}

\label{fig:ipatch}
\end{figure}

\begin{figure}
\hspace{0.05\columnwidth}\includegraphics[width=0.9\columnwidth]{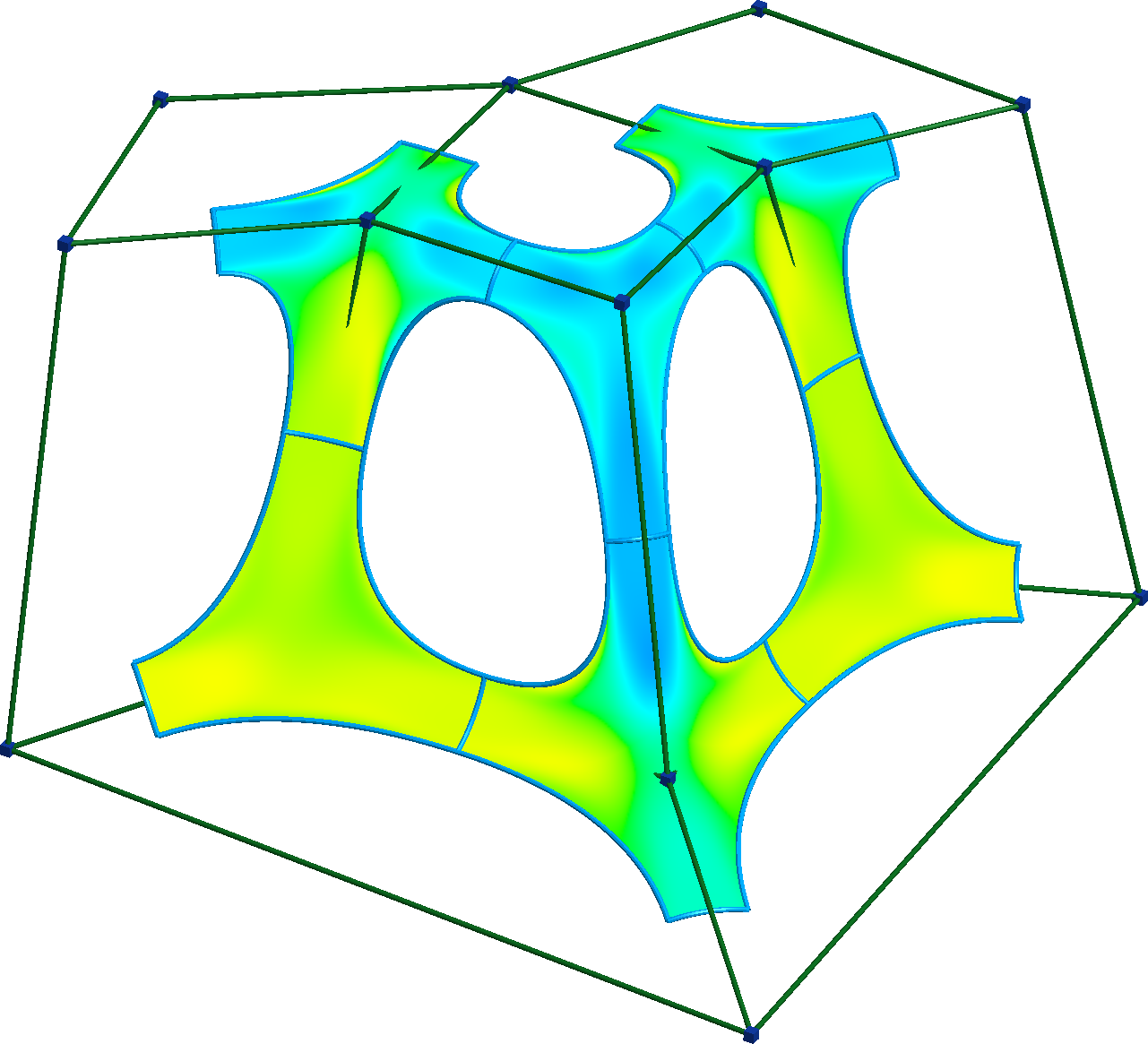}

\caption{Setback vertex blends represented by I-patches and rendered via projection.}

\label{fig:setback}
\end{figure}

We used a slightly different approach in our recent paper\cite{Sipos:2020}
where I-patches\cite{Varady:2001} are defined based on control polyhedra,
so the corner points and boundary curves of the individual patches
are known in advance. As a result, the space partitions can also be
bounded by nonplanar surfaces, and the boundary curves do not have
to be traced. The curves are then used to construct the multi-sided
$C^{0}$ Coons patch, and then we project the points onto the mesh.
The images in that paper were generated with this method. An example,
compared to Marching Cubes, is shown in Figure \ref{fig:ipatch}.
The method works even for patches defined inside concave space partitions,
for example in the case of the setback vertex blends\cite{Varady:1997}
seen in Figure \ref{fig:setback}.

\section{Conclusion\label{sec:Conclusion}}

We have proposed an algorithm for creating good quality meshes for
visualizing smooth implicit surfaces. Our method gives more accurate
results when approximating differential quantities on the surface,
due to the evenly distributed valencies of the mesh vertices. This
is an offline algorithm, and compared to Marching Cubes, it requires
significantly fewer points to achieve the same level of detail. 

However, the algorithm does not necessarily yield correct results.
It may be useful future work to provide conditions ensuring a good
tessellation.

\section*{Acknowledgements}

This project has been supported by the Hungarian Scientific Research
Fund (OTKA, No.~124727: Modeling general topology free-form surfaces
in 3D). The authors thank Tamás Várady for his valuable comments.

\bibliographystyle{unsrt}
\bibliography{paper}

\end{document}